\documentclass[aps,prb,twocolumn,superscriptaddress]{revtex4-2}
\usepackage[T1]{fontenc} 
\usepackage{float}
\usepackage{graphicx}
\usepackage{dcolumn}
\usepackage{bm}
\usepackage{color}
\usepackage{amsmath}
\usepackage{amssymb}
\usepackage{xcolor}
\usepackage{braket}
\usepackage[colorlinks = true,
            linkcolor = blue,
            urlcolor  = red,
            citecolor = blue,
            anchorcolor = blue]{hyperref}

\usepackage{mathdots}
\newcommand{\pdag}{{\phantom{\dagger}}}

\newcommand{\Tc}{T_{\text{c}} }
\newcommand{\Nc}{N_{\text{c}} }
\newcommand{\Tmf}{T^{\text{MF}}_\text{c} }
\newcommand{\cc}{\mathrm{c}}
\newcommand{\tperp}{t_{\perp}}

\newcommand{\ii}{\mathrm{i}}
\newcommand{\bk}{\mathbf{k}}

\newcommand{\kp}{k^{\prime}}

\newcommand{\cdag}{\hat{c}^{\dagger}}
\newcommand{\chat}{\hat{c}^{\pdag}}

\newcommand{\nhat}{\hat{n}}

\newcommand{\mrm}[1]{\mathrm{#1}}

\allowdisplaybreaks

\begin{document}

\title{Enhancing $\Tc$ in a composite  superconductor/metal bilayer system: a dynamical 
 cluster approximation study}
\author{Philip M. Dee}
\affiliation{Department of Physics and Astronomy, The University of Tennessee, 
Knoxville, Tennessee 37996, USA}
\affiliation{Department of Physics, University of Florida, Gainesville, Florida, 32611, USA}
\affiliation{Department of Materials Science and Engineering, University of Florida, Gainesville, Florida, 32611, USA\looseness=-1}
\author{Steven Johnston}
\affiliation{Department of Physics and Astronomy, The University of Tennessee, 
Knoxville, Tennessee 37996, USA}
\affiliation{Institute for Advanced Materials and Manufacturing, University of Tennessee, Knoxville, Tennessee 37996, USA\looseness=-1}
\author{Thomas A. Maier}
\affiliation{Computational Sciences and Engineering Division, Oak Ridge National 
Laboratory, Oak Ridge, Tennessee 37831-6102, USA\looseness=-1}
\date{\today}

\begin{abstract}
It has been proposed that the superconducting transition temperature $\Tc$ of an 
unconventional superconductor with a large pairing scale but strong phase 
fluctuations can be enhanced by coupling it to a metal. 
However, the general efficacy of this approach across different parameter regimes remains an open question. 
Using the dynamical cluster approximation, we study this question in a system 
composed of an attractive Hubbard layer in the intermediate coupling regime, where the magnitude of the attractive Coulomb interaction $|U|$ is slightly larger than the bandwidth $W$,  hybridized with a noninteracting metallic layer. 
We find that while the superconducting transition becomes more mean-field-like 
with increasing interlayer hopping, the superconducting transition temperature $\Tc$ exhibits a nonmonotonic dependence on the strength of the hybridization $\tperp$. 
This behavior arises from a reduction of the effective pairing interaction in 
the correlated layer that out-competes the growth in the intrinsic pair-field susceptibility induced by the coupling to the metallic layer. 	
We find that the largest $\Tc$ inferred here for the composite system is below the maximum value currently 
estimated for the isolated negative-$U$ Hubbard model.  

\end{abstract}

\maketitle
\section{Introduction}
%
An early and curious observation of the underdoped cuprate superconductors is that they host a 
remarkably low carrier density and correspondingly low superfluid stiffness; yet, 
they have a large pairing scale characterized by the superconducting gap and 
correspondingly short coherence length~\cite{Uemura1989,Emery1995}.
These properties give rise to a situation where Cooper 
pairing and long-range phase coherence occur at different temperatures, and the 
superconducting transition temperature $\Tc$ is significantly lower than the 
corresponding mean-field temperature scale 
$\Tmf$ ~\cite{Emery1995, Corson1999, Xu2000, Wang2005, Rourke2011, Maier2019}.
This behavior is in contrast to metallic superconductors, where pairing and 
long-range phase coherence happen simultaneously and $\Tmf=T_c$~\cite{Corson1999}. 

\begin{figure}[b]
	\includegraphics[width=1.0\columnwidth,trim=0.965cm 3.0cm 14.4cm 1.5cm,clip]{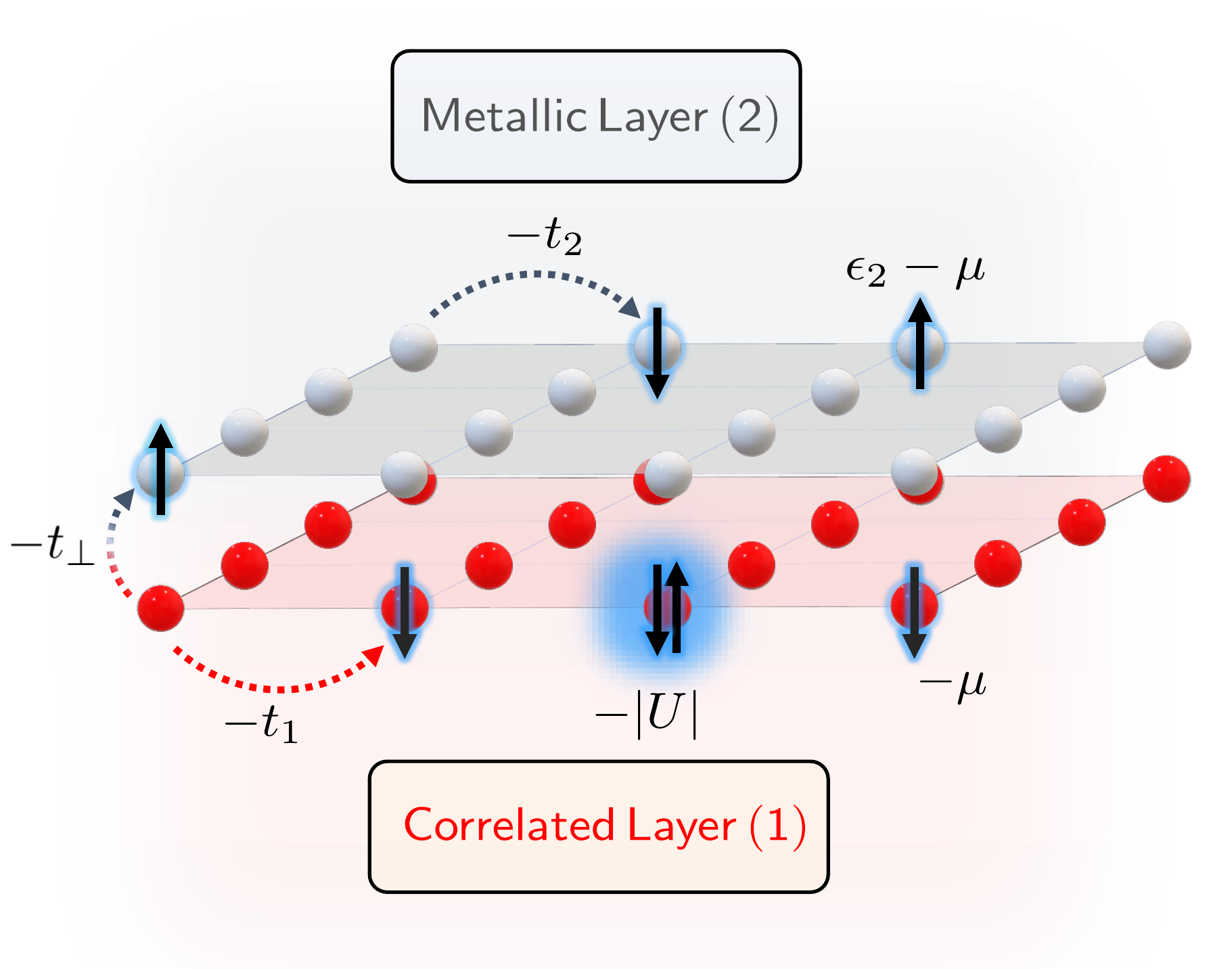}
	\caption{A cartoon depiction of the two-layer (two-orbital) model studied in this 
	work.
	Each layer consists of a square lattice. 
	The dashed arrows exemplify the possible hoppings $t_{1}$, $t_{2}$, and $t_{\perp}$, 
	and black arrows represent spin-up and down electrons. 
	Here we use $t_2=2t_1$ and $\epsilon_2=0.2t_1$, treat $\tperp$ as a variable parameter and set $-|U|=-10t_1$.
	}
	\label{fig:Bilayer_Cartoon}
\end{figure}

Kivelson~\cite{KIVELSON2002} proposed that the $\Tc$ of a superconductor 
with low superfluid stiffness could be raised closer to its mean-field value by 
coupling it to a metallic system. 
Using a disconnected (i.e., no in-plane hopping $t_1$, see 
Fig.~\ref{fig:Bilayer_Cartoon}) negative-$U$ Hubbard layer coupled to a metallic 
layer by single-particle tunneling $\tperp$, this proposal was originally 
studied perturbatively~\cite{Berg2008} for $|U|$ and $\tperp$ much smaller 
than the metallic bandwidth $W$, and later using quantum Monte Carlo (QMC) 
up to values of $|U|$ larger than the bandwidth~\cite{Wachtel2012}.
In this model, the pairing layer has zero superfluid stiffness and the 
corresponding critical temperature vanishes when the interlayer tunneling is 
switched off (i.e. $\tperp=0$). 
For small and increasing $\tperp$, phase coupling between pairing sites is enabled 
by Josephson tunneling through the metallic layer~\cite{Berg2008,Wachtel2012} and 
$\Tc$ increases; however, $\Tc$ is eventually suppressed by the same 
delocalization effects beyond intermediate values of $\tperp$.
The numerical results obtained by Watchel et al.~\cite{Wachtel2012} further 
suggest that phase fluctuations exponentially suppress $\Tc$ for small 
$\tperp$. 
Moreover, they found that the highest $\Tc$ that can be achieved by varying 
$\tperp$ were quite modest. For example, $\Tc$ is three to four times smaller 
than $\Tmf$ for small and intermediate $|U|/t$, where $t$ is the hopping amplitude in the metallic layer ($t_2$ in our model).  Moreover, $T_c$ remains below 
the highest $\Tc$ in the isolated 2D negative-$U$ Hubbard model for large $|U|/t$.  

The intermediate coupling regime $|U|\sim W$, where $W$ is the bandwidth in the 
negative-$U$ layer, but with intraplane hopping in the negative-$U$ layer 
restored was later addressed using QMC~\cite{Zujev_2014}.
In that case, the correlated layer has a small but nonzero superfluid stiffness, 
even when $\tperp=0$, and the superconducting transition follows the 
Berezinskii-Kosterlitz-Thouless (BKT) universality class of the XY model. 
For increasing $\tperp$, Ref.~\citenum{Zujev_2014} observed a proximity 
effect-induced suppression of the pairing correlations in the correlated layer, 
while the metallic layer exhibited nonmonotonic behavior with a maximum in the 
pairing correlations at intermediate $\tperp$. 
In other words, pairing is layer (or orbital) dependent for small $\tperp$, an 
observation that is reminiscent of the orbital selective behaviors seen in 
many multi-orbital Hubbard models~\cite{Werner2007,Tocchio2016}. 
Eventually, the pairing in both layers is suppressed simultaneously beyond a 
critical value of $\tperp$. 

While the work in Ref.~\onlinecite{Zujev_2014} provided a finite-size analysis 
of the pairing correlations above $\Tc$, it did not determine $\Tc$ as a function 
of $\tperp$. 
Estimating the latter is crucial, however, because the strength of the pairing 
correlations at $T>\Tc$ can be a poor indicator for the actual $\Tc$ realized 
in a system~\cite{Mai2021, Karakuzu2021}. 
We address this issue by studying a negative-$U$ Hubbard model coupled to a 
noninteracting layer using a quantum Monte Carlo (QMC) dynamic cluster approximation (DCA). 
Here, we focus on the nature of the superconducting transition and determining 
$T_c$ as a function of $\tperp$ by solving the Bethe-Salpeter equation in the 
particle-particle channel. 
Since the DCA incorporates long range physics in the thermodynamic limit 
through a coarse-graining of momentum space, it also provides a different 
numerical perspective than finite cluster QMC methods. 
Our results show that for large clusters, $\Tc$ is enhanced for finite $\tperp$ 
beyond the system's $\Tc$ when $\tperp=0$.
That is, an increased single-particle tunneling between the layers reduces the 
effects of phase fluctuations present in negative-$U$ layer and thus leads to 
an increased $\Tc$.
We also discuss how the increased interlayer coupling leads to an increase in the intrinsic pair-field susceptibility, which competes with a reduction in the 
effective pairing interaction, leading to a nonmonotonic dependence of $\Tc$ on $\tperp$. 
These two quantities suggest signs of crossover behavior similar to the BEC-BCS 
crossover found in the pure negative-$U$ Hubbard 
model~\cite{Scalettar1989, Keller2001, Kaneko2014, Hazra2019}. 


\section{Model and Methods}

\subsection{Two-component Model}
%
Our composite system consists of a correlated negative-$U$ Hubbard layer and a 
noninteracting metallic layer connected through interlayer single-particle 
tunneling $\tperp$ (see Fig.~\ref{fig:Bilayer_Cartoon}). 
Both layers have square lattice geometry with identical lattice spacing.  
The associated bilayer Hamiltonian is defined as
\vspace{1cm}
\begin{multline}\label{eqn:H}
\hat{H} 
	=  -\sum_{\langle ij\rangle, l, \sigma} t^\pdag_{l}
	(\cdag_{il\sigma} \chat_{jl\sigma} + \text{H.c})
	- |U| \sum_{i} \nhat_{i1\uparrow} \nhat_{i1\downarrow}  \\
	+\sum_{i,l,\sigma} (\epsilon_{\text{2}}\delta_{l 2} - \mu)
	\nhat_{il\sigma}
	-t^\pdag_{\perp} \sum_{i,\sigma}(\cdag_{i1\sigma}\chat_{i2\sigma}+\text{H.c.}),
\end{multline} 
where $ \hat{c}_{il\sigma}^{\dagger} $ ($ \hat{c}_{il\sigma}^{\pdag}$) 
creates (destroys) an electron on the $ i $\textsuperscript{th} site of the 
$l=1$ or 2 layer with spin $ \sigma\,(=\uparrow,\downarrow) $ and 
$\nhat^\pdag_{il\sigma} = 
\hat{c}_{il\sigma}^{\dagger} \hat{c}_{il\sigma}^{\pdag}$.
The in-plane nearest-neighbor hoppings $t_{l}$ are fixed such that 
$t_{1}\equiv t=1$ and $t_{2}=2t$, whereas the interlayer tunneling $\tperp$ is 
a variable parameter.     
An attractive on-site Coulomb interaction $-|U|$ in the correlated layer 
$(l=1)$ is responsible for the formation of local ($s$-wave) Cooper pairs. 
Lastly, the noninteracting metallic layer $(l=2)$ has an additional on-site 
energy term that is used to shift the van Hove singularity slightly above the 
chemical potential $\mu$ (we set $\epsilon_{2}-\mu = 0.2t$).
%

\subsection{Methods}
%
We studied the Hamiltonian in Eq.~(\ref{eqn:H}) using the DCA++ 
code~\cite{HAHNER2020}, a state-of-the-art implementation of the DCA method.
In this formalism, the lattice problem is reduced to a finite-size cluster 
embedded in a mean-field that is self-consistently determined to represent the 
system beyond the cluster~\cite{MaierRMP2005}.
Our model is an effective two orbital model resulting in a $2N_{\cc}$-site 
cluster problem for an in-plane cluster of size $N_c$, which is solved using a 
continuous-time auxiliary-field QMC algorithm~\cite{Gull_2008,Gull_2011,Gull_2011_RMP}.
In this work, we examine several different in-plane cluster sizes including 
4$\times$4 ($N_{\cc}=16$), 6$\times$6 ($N_{\cc}=36$), 8$\times$8 ($N_{\cc}=64$), 
and 10$\times$10 ($N_{\cc}=100$). 
%

The QMC simulations utilized 6000 independent Markov chains to collect 
$2\times10^{6}$ to $5\times10^{6}$ total measurements. 
The model with negative-$U$ interaction does not have a sign problem but tends 
to produce long autocorrelation times. 
To combat the latter, each of the contributing measurements was made after 
skipping 50-100 Monte Carlo sweeps to further ensure statistically independent 
sampling. 
A typical DCA calculation for our model (i.e., for a single set of model 
parameters) converges in 6-8 iterations depending on the cluster size.

Throughout this work, we allow the chemical potential $\mu$ to vary such that 
the filling in the correlated layer remains fixed at 
$n_{1} \equiv \langle \nhat_{i1} \rangle = 0.75$, where 
$\nhat_{i1}=\nhat_{i1\uparrow}+\nhat_{i1\downarrow}$. 
The filling of the metallic layer is allowed to take whatever value is 
necessary to satisfy thermodynamic equilibrium as a result. 
This choice of filling avoids any complications that might stem from a perfectly 
nested Fermi surface as seen at half-filling and it still gives us access to 
the superconducting transition in the negative-$U$ model.

For the isolated ($\tperp=0$) 2D negative-$U$ Hubbard model away from 
half-filling, the system has an $s$-wave superconducting ground 
state~\cite{Moreo1991, Scalettar1989, Keller2001, Paiva2010, Kaneko2014, Hazra2019, Fontenele2022}.
When $|U|/t \ll 1$, the system adopts a weak coupling BCS state and eventually 
crosses over to a Bose-Einstein condensate (BEC) of hard-core on-site bosons 
for $|U|/t \gg 1$~\cite{Scalettar1989,Keller2001,Kaneko2014,Hazra2019}.
Consequently, $\Tc$ is a nonmonotonic 
function of $|U|/t$ that peaks at intermediate $|U|/t\approx4-6$, and gradually 
tapers off 
in the presence of increasingly stronger phase fluctuations at larger values of 
$|U|/t$~\cite{Paiva2010, Kaneko2014, Fontenele2022}.
We are interested in the question of whether the reduction in $\Tc$ due to phase 
fluctuations can be reversed in the composite system. 
We therefore set $-|U|=-10t$ in the correlated layer and vary the interlayer hopping $\tperp$ to study its effects on the $\Tc$ of the composite system. 
%

We estimate $\Tc$ by solving the Bethe-Salpeter 
equation \cite{Maier2006, Maier2006b} as an eigenvalue problem: 
\begin{equation}\label{eqn:BSE}
 	-\frac{T}{N_{\cc}}\sum_{\kp}\Gamma^{\mrm{pp}}(k,\kp)G(\kp)G(-\kp)
 	\phi_{\alpha}(\kp) = \lambda_{\alpha}\phi_{\alpha}(k)\,.
\end{equation}
Here, the eigenvalues and corresponding eigenvectors are given by 
$\lambda_{\alpha}$ and $\phi_{\alpha}$, respectively; $G(k)$ is the dressed 
single-particle propagator and $\Gamma^{\mrm{pp}}(k,\kp)$ the irreducible 
particle-particle vertex, both obtained from the DCA and written compactly 
using the notation $k\equiv (\bk,\ii\omega_{n})$, where $\bk$ is the 
momentum and $\ii\omega_{n}$ is a fermionic Matsubara frequency 
$\omega_{n}=(2n+1)\pi T$. 
The index $\alpha$ ranges over the entire set of eigensolutions, but we will limit 
our discussion to the solution corresponding to the largest eigenvalue, denoted 
$\lambda_{s}$. 
A superconducting transition occurs when the leading eigenvalue 
$\lambda_{s}(\Tc)=1$. 
In our case, we find that the leading eigenvector has $s$-wave symmetry for 
all the values of $\tperp$ we consider.  
%

\section{Results}
\subsection{Temperature dependence of the pairing correlations}
\begin{figure}[t]
	\includegraphics[width=1.0\columnwidth]{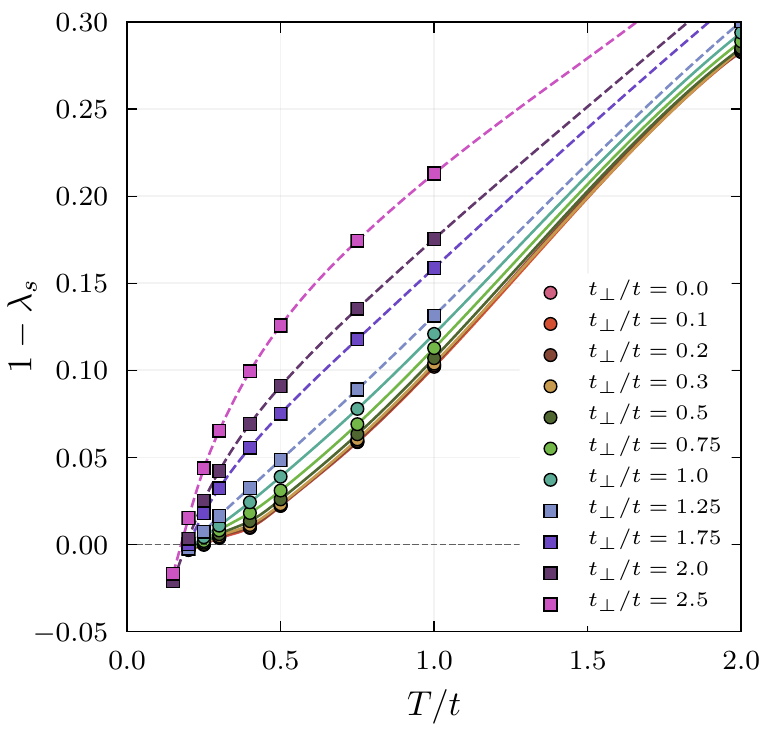}
	\caption{Temperature dependence of $1-\lambda_{s}(T)$ for the composite 
	bilayer system on a $N_{\cc}=16$ site cluster for several different values 
	of the interlayer hopping in the interval $\tperp/t\in[0,2.5]$.
	Filled circles (squares) with solid (dashed) lines depict results for 
	$\tperp/t\leq 1$ ($\tperp/t > 1$). }
	\label{fig:one_minus_lam}
\end{figure}

We first examine the temperature dependence of $1-\lambda_{s}(T)$ for several 
values of $\tperp$, plotted in Fig.~\ref{fig:one_minus_lam} for $\Nc = 16$.
Starting from high-temperature ($T/t=2$), we cool the composite system down to 
$T\lesssim  \Tc$, which is identified by the temperature at which 
$1-\lambda_{s}(T=\Tc)=0$. 
The family of curves plotted in Fig.~\ref{fig:one_minus_lam} represent 
different values of the interlayer hopping between 0 and $2.5t$ but with 
all other model parameters identical (except $\mu$, which varies as to 
fix $n_{1}=0.75$).
All curves for $\tperp/t\leq 1$ are denoted by filled circles and solid lines 
(to guide the eye) and the remaining curves with $\tperp/t > 1$ are depicted 
with filled squares and dashed lines. 

%

%

As noted, the superconducting transition in this model is expected to follow the 
BKT universality class. 
Since the DCA embeds the finite-size cluster in a mean-field, the calculated 
temperature dependence will cross over to mean-field behavior when the correlation 
length exceeds the cluster size~\cite{MaierRMP2005}. 
But at higher temperatures, when the correlations are still contained 
within the cluster, the DCA results will exhibit the true temperature dependence 
of the system in the thermodynamic limit.
For small $\tperp/t < 1$, the curves display convex behavior, indicating the 
presence of phase fluctuations and BKT behavior~\cite{Maier2019}.
For larger $\tperp/t > 1$, the temperature dependence of $1-\lambda_{s}(T)$ 
changes qualitatively to a more BCS-like behavior, exhibiting logarithmic 
$\ln(T/\Tc^{\text{MF}})$ dependence. 
The change in curvature around $\tperp = 1$ is similar to what is observed for the 
repulsive Hubbard model with increasing hole doping~\cite{Maier2019}.
It reflects a change in the nature of the superconducting phase transition and 
the decreasing strength of phase fluctuations as $\tperp$ increases. 
The results in Fig.~\ref{fig:one_minus_lam} therefore suggest that coupling to 
the metallic layer makes the superconducting transition more mean-field-like. 


\begin{figure}[t]
	\includegraphics[width=0.9\columnwidth]{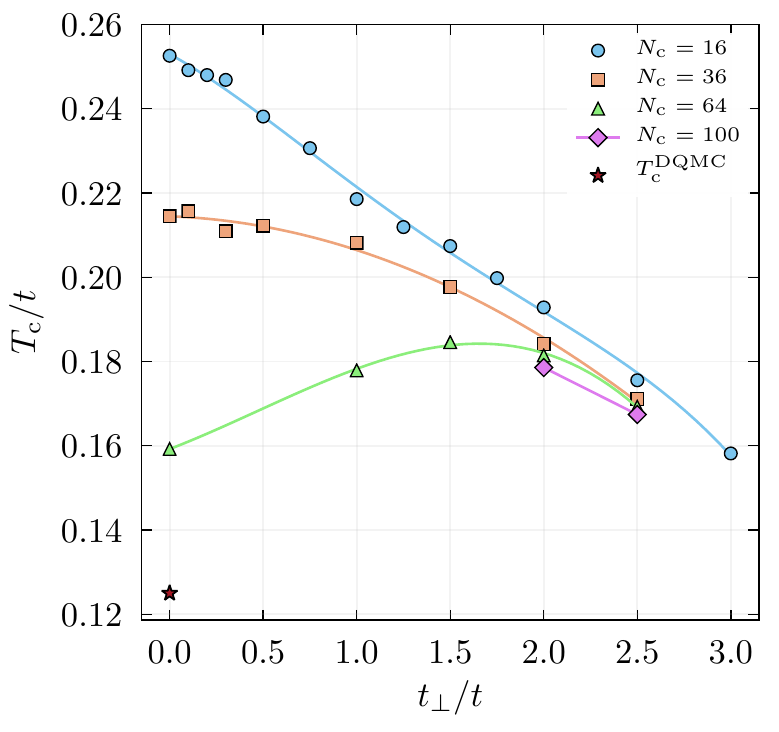}
	\caption{Superconducting critical temperature $\Tc/t$ as function of 
	interlayer tunneling $\tperp/t$ for $\Nc=16$, 36, 64 and 100.
	For $\Nc = 16 $ and 36, $\Tc$ is a monotonically decreasing function 
	of $\tperp$, but with strong indications of a strong cluster size dependence 
	for small $\tperp$. 
	However, when $\Nc = 64 $, $\Tc$ is nonmonotonic as a function 
	of $\tperp$, suggesting that large cluster sizes are required to 
	capture the effects of phase fluctuations on $\Tc$ for small $\tperp$. 
	We include two points for $\Nc = 100$ to demonstrate that $\Tc$ is well captured 
	by smaller clusters when $\tperp / t > 2$.
	The red star indicates the estimate for $\Tc(\tperp/t=0)$ as $\Nc\rightarrow\infty$ stemming from a recent DQMC study 
	(Ref.~\cite{Fontenele2022}) of the 2D negative-$U$ model.
	The solid lines are intended to guide the eye. 
	\label{fig:Tc_vs_tperp} }
\end{figure}

\subsection{Transition temperature vs. interlayer coupling}
Fig.~\ref{fig:Tc_vs_tperp} plots the estimated $\Tc$ values for  $\Nc=16$, 36, 
64, and 100. 
Unfortunately, the long autocorrelation times affecting the small $\tperp$ 
calculations prevent us from obtaining $\Tc$ estimates for larger clusters. 
Moreover, we observe significant cluster size dependence in the extracted 
values of $\Tc$, particularly when $\tperp/t < 1$. 
This occurs because larger clusters are needed to account for the long-range 
spatial fluctuations that suppress $\Tc$ in this regime.
(This observation agrees with recent DQMC studies, which found that large 
clusters were needed to accurately estimate $T_c$ for the negative-$U$ Hubbard model~\cite{Paiva2010,Fontenele2022}.) 
Despite these limitations, we are able to draw some general conclusions about 
the pairing tendencies in the model, which we now address.

The results from the two smallest cluster sizes ($\Nc=16$, 36) are qualitatively 
similar in that the coupling $\tperp$ to the metallic layer suppresses $\Tc$ in 
this parameter regime. 
For $\Nc=64$, however, $\Tc(\tperp)$ displays nonmonotonic behavior with a 
maximum $\Tc$ near $\tperp/t = 1.5$. Interestingly, the maximum occurs near the crossover between the BEC and BCS behavior observed in the temperature dependence of $\lambda_s(T)$ in Fig.~\ref{fig:one_minus_lam}. We will further discuss this point in Sec.~D below.
This result suggests that $\Tc$ can indeed be optimized by adjusting the 
interlayer coupling; however, to determine the precise magnitude of this 
enhancement, we must contend with the finite-size effects in the small 
$\tperp$ regime. 

The most recent estimates for the negative-$U$ Hubbard model~\cite{Fontenele2022}, 
based on DQMC and extrapolated to the thermodynamic limit, place $\Tc/t \approx 
0.12$ for our model parameters, as indicated in Fig.~\ref{fig:Tc_vs_tperp}.
We therefore expect the estimated $\Tc$ from DCA to continue to decrease for larger 
cluster sizes ($\Nc>64$) and $\tperp/t\ll 1$, until it is on par with this value. 
Conversely, we do not observe such a strong cluster size dependence for larger 
$\tperp$ values. 
For example, the results for $\Nc=100$ and $\tperp \geq 2$ almost lay on top of 
the corresponding $\Nc=64$ data points. 
We therefore believe that our results are relatively well converged for larger  
$\tperp/t$. 
Combined, the results in Fig.~\ref{fig:Tc_vs_tperp} then indicate that the 
$\Tc$ of the composite system is indeed larger than the $\tperp = 0$ case in 
the $|U|\gtrsim W$ regime, with the largest $\Tc$ values occurring for 
$\tperp \approx 1.5t-2t$. 
Moreover, given the rate of convergence observed in Fig.~\ref{fig:Tc_vs_tperp}, 
it is clear that the maximum value of $\Tc(\tperp)$ is comparable to the 
maximum $\Tc \approx 0.16t$ value obtained in the negative-$U$ model~\cite{Fontenele2022}. 
This $\Tc$ value is relatively constant across a range of $|U|/t$ from 4-6 and 
electron filling $\langle \hat{n}\rangle$ between 0.70 and 0.88.

\subsection{Effective pairing interaction and pair mobility}
Thus far, we have demonstrated that $\Tc$ follows a nonmonotonic dependence 
on $\tperp$ when the DCA cluster size is sufficiently large.
We now turn to the question of what drives this nonmonotonicity by examining the 
$t_\perp$ dependence of the effective pairing interaction and the intrinsic pairfield susceptibility, both of which determine the leading eigenvalue $\lambda_s$ of the 
Bethe-Salpeter equation. 

The $s$-wave pair field susceptibility $P_s$ is given by
\begin{equation} \label{eq:Ps}
    P_s(T) = \int_0^\beta \mathrm{d}\tau \,\langle T_\tau \hat{\Delta}(\tau)\hat{\Delta}^{\dagger}(0)\rangle
\end{equation}
with $\hat{\Delta}^\dagger = \frac{1}{N}\sum_{i,l} \cdag_{il\uparrow}\cdag_{il\downarrow}$. Its leading order term
\begin{equation}\label{eqn:Ps0_new}
	P_{s0}(T) = \frac{T}{N}\sum_{\substack{k \\ l_{1},l_{2},l_{3}, l_{4}}} G^{l_{1}l_{3}}(k)G^{l_{2}l_{4}}(-k)
\end{equation} 
defines the intrinsic pairfield susceptibility $P_{s0}(T)$. With these two quantities, we can then define an effective pairing interaction $V_s$ through
\begin{equation}
    V_{s}(T) = P_{s0}^{-1}(T) - P_{s}^{-1}(T)\,.
\end{equation}
We note that using this definition for $V_s$, we find that the product $V_s P_{s0}$ gives values very similar to those for the leading eigenvalue $\lambda_s$ with a difference of at most 5\%. 

\begin{figure}[t]
	\includegraphics[width=1.0\columnwidth]{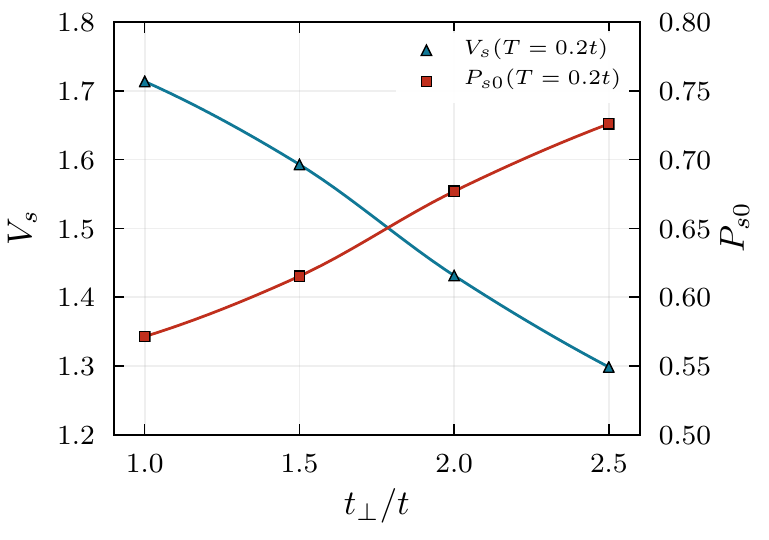}
	\caption{Effective interaction $V_{s}(\tperp)$ and irreducible pair-field 
	susceptibility $P_{s0}(\tperp)$ evaluated at $T/t = 0.2t$ and $q=0$. 
	Results were obtained on a cluster size of $\Nc=64$. 
	The solid lines are included to guide the eye.
	}
	\label{fig:Vs_Ps0_vs_tperp}
\end{figure}
Figure~\ref{fig:Vs_Ps0_vs_tperp} plots  $V_{s}$ and $P_{s0}$ at a fixed 
temperature $T/t=0.2$ across the range of $\tperp/t$ spanning the 
$\Tc$ ``dome'' in Fig.~\ref{fig:Tc_vs_tperp}.
As the interlayer tunneling $\tperp$ increases, the effective pairing interaction $V_s$ 
\emph{decreases} monotonically. 
Conversely, the intrinsic pair-field susceptibility \emph{increases} as the 
coupling to the metallic layer grows. 
The enhancement of $P_{s0}$ would raise $\lambda_s$ and therefore $\Tc$, but it competes with the  
decrease in in the pairing interaction, leading to a nonmonotonic dependence of $\Tc$ on $\tperp$. 
Since $P_{s0}(\tperp)$ is a measure of the states available to form $s$-wave pairs, it provides an indirect measure of the superfluid phase stiffness.  
We can, therefore, conclude that the increase in $\Tc$ is indeed being driven 
by enhanced superfluid phase stiffness but that this is ultimately 
counteracted by a decrease in the effective pairing interaction. 

We note that $V_{s}$ and $P_{s0}$ have been defined here 
using contributions from the entire system, i.e. the sum over $l$ in Eqs.~(\ref{eq:Ps}) and (\ref{eqn:Ps0_new}) runs over both layers.
Although not shown, we have repeated the analysis above but restricting 
the sums separately over just the correlated or the metallic layer.
In this case, the behavior of $V_{s}$ and $P_{s0}$ for the correlated layer 
is qualitatively identical to that shown in Fig.~\ref{fig:Vs_Ps0_vs_tperp}; however, for the metallic layer, the trends are swapped with $V_{s}$ 
increasing and $P_{s0}$ decreasing.
%

\subsection{Discussion}
While our DCA calculations provide direct access to the thermodynamic 
limit, they are also considerably more expensive than finite cluster determinant 
QMC (DQMC) calculations. 
Because of this, a shortcoming of our analysis is that it considers a narrow 
range of model parameters. 
However, a survey of the most relevant results from previous studies 
suggests that many of the qualitative trends we observe are likely insensitive 
to minor variations in model parameters.
For instance, Ref.~\cite{Zujev_2014} examined other filling factors 
($n_{1}=0.6, 0.8, 1.0$) and attractive interaction strengths 
($U/t=-4, -6, -10$). 
The authors found that smaller values of $n_{1}$ are only 
marginally better for inducing pairing correlations in the metallic layer and 
that less negative values of $U$ decrease pairing significantly in both layers.  
Thus, we expect our results to be representative of the regime where the size 
of the interaction and the bandwidth are comparable and pairing is more 
significant. 
Although a direct comparison with Ref.~\cite{Zujev_2014} is not possible, 
their calculations (for similar parameters) show that the static pair structure 
factor decreases monotonically with increasing $\tperp$ in the correlated 
layer and peaks at finite $\tperp$ in the metallic layer.
However, the strength of the induced correlations in the metallic layer 
are small compared with the correlated layer and not representative of a 
superconducting transition over the temperatures they studied.
It therefore may be necessary to compare our results directly to those 
from a method like DQMC in the future.
There, we could better gauge the role of the mean-field in the DCA method 
in situations where phase fluctuations are strong. 

The rise and fall of $\Tc$ with $\tperp$ suggests that this composite 
bilayer model provides a route to enhancing $\Tc$ as discussed 
in Refs.~\cite{KIVELSON2002} and  \cite{Berg2008}, even for this 
nonperturbative case where $t_{1}\neq 0$ and $|U|\gtrsim W$.
However, the $\Tc$ enhancement we observe is modest relative to the cases 
examined in previous works~\cite{Berg2008,Wachtel2012}, which focused 
on models where the correlated layer has virtually no superfluid stiffness 
when $\tperp=0$. 
Our model has small  phase stiffness in the correlated layer by 
tuning the interaction into a regime of increasing phase fluctuations and 
allowing comparable hopping amplitudes in both layers. 
These results suggest that the details of both the correlated and metallic 
layers can play a crucial role in determining the $\Tc$ values ultimately 
achieved in a composite system. 
While this result calls for a more exhaustive study of the model space, it 
also indicates that opportunities for additional engineering of the layers 
exist. 
%

The evolution of the superconducting transition from KT-like to BCS-like is 
reminiscent of the BCS-BEC crossover~\cite{Leggett1980,Ries2015} discussed in 
the context of the 2D negative-$U$ Hubbard model~\cite{Scalettar1989,Keller2001}. 
In the latter scenario, one goes from BEC to BCS superconductivity by 
systematically lowering $|U|$ from the intermediate-strong coupling regime to 
the weak coupling regime.
Importantly, $\Tc(|U|)$ is found to have a maximum for $|U|\sim W$.  
In the bilayer model, although $|U|$ is kept constant, we find that the pairing 
interaction is effectively reduced through an increase in the interlayer 
tunneling amplitude $t_{\perp}$. 
Interestingly, like the purely 2D case, we find that $\Tc$ in the composite 
system also has a maximum with decreasing pairing interaction, i.e. when $\tperp$ is increased, near  the crossover between the BEC and BCS regimes. Based on this observation, one may speculate that the results for the composite system can be rationalized in terms of an effective single-layer negative-$U$ model, in which the hopping amplitude has been enhanced effectively by the hopping through the metallic layer. Additional calculations are needed to determine to what extent this is indeed the case and will be the subject of future studies.
%

\section{Summary \& Conclusion}
We have examined a composite negative-$U$ Hubbard/noninteracting metallic bilayer 
system using DCA-QMC calculations to study the relationship between the 
superconducting $\Tc$ and the interlayer single-particle tunneling. 
Our work expands on previous studies~\cite{Berg2008,Wachtel2012} by focusing on 
a regime where the magnitude of the attractive interaction is comparable to the 
bandwidth ($|U| \gtrsim W$), and \textit{both} layers have finite bandwidth (i.e. 
nonzero intralayer hopping). 
Moreover, we complement Ref.~\cite{Zujev_2014} by estimating  
$\Tc$ directly and computing the system's effective pairing interaction and 
superfluid stiffness as a function of $\tperp$. 
%

We found that $\Tc$ displays nonmonotonic behavior and reaches a maximum at a finite value of the interlayer 
tunneling $\tperp$, a trend that emerges when the DCA cluster size becomes sufficiently 
large (i.e., when $\Nc\gtrsim 64$) to capture the necessary spatial fluctuations. 
For smaller clusters, phase fluctuations are suppressed by the 
mean-field and the $\Tc$ is overestimated, especially for small tunneling 
values. 
The effective pairing interaction in the correlated layer decreases 
monotonically with increasing $\tperp$, thereby lowering the pairing scale. 
However, we see a competing increase in the irreducible pair-field 
susceptibility up to a finite value of $\tperp$, which acts to increase 
$\Tc$ over the same range.
Our results suggest that the peak $\Tc$ may correspond to a crossover between 
tightly formed BEC pairs and longer range BCS pairs, much like in the 
negative-$U$ Hubbard model. 

For small interlayer tunneling, the superconducting transition displays signs of strong phase fluctuations. 
As the interlayer  tunneling increases, we observe a shift toward 
a BCS-like logarithmic temperature dependence. 
Interestingly, this confirms that the superconducting transition in the 
composite system does inherit a more mean-field-like character through the 
interlayer hybridization. 
However, this partial recapture of the mean-field pairing scale produces 
only a modest enhancement of $\Tc$ relative to the isolated layer. 
We speculate that this enhancement could be further increased by considering 
metallic layers that can retain some degree of the large pairing interaction. 
For example, coupling to a metal with strong electron-phonon coupling could help 
counteract the reduction in $V_s$. 
\\
\\  
The Department of Energy will provide public access to these results of federally sponsored research in accordance with the DOE Public Access Plan (\url{http://energy.gov/downloads/doe-public-access-plan}.).

{\it Acknowledgments} ---  
The authors thank D.~J. Scalapino and D. Orgad for useful comments on the 
manuscript. S.~.J. and T.~A.~M. were supported by the Scientific Discovery 
through Advanced Computing (SciDAC) program funded by the U.S. Department of 
Energy, Office of Science, Advanced Scientific Computing Research and Basic 
Energy Sciences, Division of Materials Sciences and Engineering.  
P.~D. was supported by the U.S. Department of Energy, Office of Science, Office 
of Workforce Development for Teachers and Scientists, Office of Science Graduate 
Student Research (SCGSR) program. 
The SCGSR program is administered by the Oak Ridge Institute for Science and 
Education for the DOE under contract number DE‐SC0014664.
P.~D. also acknowledges support from the U.S. Department of Energy, Office of 
Science, Office of Basic Energy Sciences, under Award Number DE-SC-0020385
while writing this paper. T. A. M. acknowledges additional support from the U.S. Department of Energy (DOE), Office of Science, Basic Energy Sciences (BES), Materials Sciences and Engineering Division for analyzing some of the results and writing the paper. 
This research used resources of the Oak Ridge Leadership Computing Facility, 
which is a DOE Office of Science User Facility supported under Contract 
DE-AC05-00OR22725.
This manuscript has been authored by UT-Battelle, LLC, under Contract No. DE-AC0500OR22725 with the U.S. Department of Energy. The United States Government retains and the publisher, by accepting the article for publication, acknowledges that the United States Government retains a non-exclusive, paid-up, irrevocable, world-wide license to publish or reproduce the published form of this manuscript, or allow others to do so, for the United States Government purposes. 

\bibliography{Ref_file}
\end{document}